\documentclass[doublecol,figures]{epl2}

\usepackage{amssymb}

\title{Na ordering imprints a metallic kagom\'{e} lattice onto the Co
planes of Na$_{2/3}$CoO$_{2}$}%
\shorttitle{Na ordering imprints a metallic kagom\'{e} lattice onto the Co
planes of Na$_{2/3}$CoO$_{2}$}

\author{H.~Alloul\inst{1}\thanks{E-mail: \email{alloul@lps.u-psud.fr}}
\and I.R.~Mukhamedshin\inst{1,2} \and T.A.~Platova\inst{1,2} \and
A.V.~Dooglav\inst{2}}%
\shortauthor{H.~Alloul \etal}

\institute{
  \inst{1} Laboratoie de Physique des Solides, UMR CNRS 8502, Univ. Paris-Sud, 91405 Orsay, France\\
  \inst{2} Physics Department, Kazan State University, 420008 Kazan, Russia\\
}

\pacs{71.28.+d}{Narrow-band systems; intermediate-valence solids}%
\pacs{76.60.-k}{Nuclear magnetic resonance and relaxation}%
\pacs{71.27.+a}{Strongly correlated electron systems; heavy fermions}%

\abstract{ We report $^{23}$Na and $^{59}$Co nuclear magnetic (NMR) and
quadrupolar resonance (NQR) studies for the $x=2/3$ phase of the lamellar
oxide Na$_{x}$CoO$_{2}$, which allowed us to establish reliably the atomic
order of the Na layers and their stacking between the CoO$_{2}$ slabs. We
evidence that the Na$^{+}$ order stabilizes filled non magnetic Co$^{3+}$
ions on 25\% of the cobalt sites arranged in a triangular sublattice. The
transferred holes are delocalized on the 75\% complementary cobalt sites
which unexpectedly display a planar cobalt kagom\'{e} structure. These
experimental results resolve a puzzling issue by precluding localized
moments pictures for the magnetic properties. They establish that the quasi
ferromagnetic properties result from a narrow band connecting a frustrated
arrangement of atomic orbitals, and open the route to unravel through
similar studies the electronic properties of the diverse ordered phases of
sodium cobaltates. }

\begin{document}

\maketitle

\section{Introduction}

Peculiar atomic structures such as chains, ladders or the graphene honeycomb
often exhibit remarkable singular physical properties, as does the
triangular cobalt network in Na cobaltates \cite{Fouassier} which displays
high thermopower \cite{Terasaki} and superconductivity \cite{Takada}. There,
the control of carrier content of the CoO$_{2}$ planes by varying Na
concentration between the planes yields a totally counter-intuitive sequence
of magnetic properties including anomalous paramagnetism \cite{Foo}, charge
disproportionation \cite{CoPaper}, metallic antiferromagnetism \cite%
{Bayrakci,Helme}.

An original aspect of these systems is that the Co atoms are stacked between
two triangular oxygen layers with distorted CoO$_{6}$ octahedrons, so that a
large crystal field stabilizes the Co ions in a low spin state by lifting
the degeneracy of the cobalt 3$d$ levels. In a Co$^{3+}$ configuration the
six lower energy levels ($t_{2g}$) are filled, with a total spin $S=0$,
while Co$^{4+}$ should only retain one hole in the $t_{2g}$ multiplet, with $%
S=1/2$, so that original magnetic properties have been anticipated to result
from these local spins associated with charge ordering \cite{Baskaran}. This
charge order intrinsic to the Co planes would depend of $x$ and would yield
specific metallic and magnetic properties. For instance, for $x=2/3$, the 2D
charge ordered state would be a honeycomb network of Co$^{3+} $ ions
intermixed with a triangular lattice of Co$^{4+}$.

We formerly established experimentally by NMR \cite{CoPaper} that such a
charge order does not occur, but that the charges disproportionates between
Co$^{3+}$ and cobalt sites with an average formal valence of about 3.5$^{+}$%
. The differentiation of three Na sites \cite{NaPaper}, in a phase with $%
x\sim 0.7$ (redefined as $x=0.67$ or H67 in ref.~\cite{ourEPL2008}) implied
an associated atomic Na ordering with a unit cell larger than that of the
honeycomb lattice, in view of the large number of cobalt sites detected by
NMR \cite{CoPaper}. This further differentiates the Na cobaltates from the
cuprates for which copper has a uniform charge while dopants are usually
disordered, which  influences the physical properties \cite{FRA-alloul}.
Indeed the Na atomic orderings stabilized for specific $x$ values are found
to play a role in determining the ground state metallic and magnetic
properties \cite{ourEPL2008}. But, while Na ordering has been evidenced by
various experiments \cite{NaPaper,Zandbergen,Roger}, the actual atomic order
and its incidence on the local scale electronic properties is still unclear
\cite{ourEPL2008}, except for two limiting cases, the $x=1$ filled band
insulator and the $x=1/2$ ordered "chain-like" compound \cite%
{Bobroff,Yokoi,Mendels}.

Let us point out that the cobalt plane physical properties are influenced by
the Na order in the two layers between which the CoO$_{2}$ slab is
sandwiched, so that any connection between structure and physical properties
\textit{requires the knowledge of the 3D structure}. This is in principle
possible using diffraction techniques and 3D order has indeed been seen by
neutron scattering \cite{Roger}, X rays on single crystals \cite{Shu} and
even in our powder samples \cite{ourEPL2008,Lang2}. However, so far,
materials complications linked with phase mixing, difficult accurate
determination of Na content and existence of twins in the Na order have
prevented the finalization of such studies.

It is well known that NMR is an ideal technique to link structural and
electronic properties, as it allows to measure local magnetic properties and
is also sensitive to structural properties. Indeed, nuclear spins with $%
I>1/2 $ sense the local distribution of charges through the electrostatic
interaction which couples the quadrupolar moment $eQ$ of the nuclear charge
with the electric field gradient (EFG) at the nuclear site \cite{Slichter}.
In view of the complicated NMR spectra \cite{CoPaper}, we demonstrate here
that the structural and magnetic properties are better sorted out if one
uses altogether NQR, that is a direct determination of the zero field
splitting of the nuclear spin levels \cite{Slichter} due to the quadrupole
interaction. We shall demonstrate below that this approach applied here to
the $x=2/3$ phase allows us to evidence unambiguously that the Na ordering
drives an in plane metallic electronic kagom\'{e} organization \cite{Mekata}%
. Efforts to synthesize such geometric structures at the atomic level have
been successful recently on producing a remarkable S=1/2 \textit{insulating}
spin liquid state \cite{Shores,Mendels3}, but have not allowed so far to
synthesize metallic states.

\section{ Differentiation of sites by NQR and NMR}

\begin{figure}[tbp]
\onefigure[width=1\linewidth]{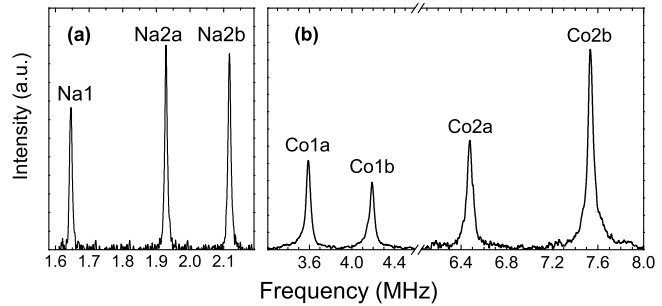}
\caption{NQR signals of the various $^{23}$Na and $^{59}$Co sites in Na$%
_{2/3}$CoO$_{2}$. (a) For $^{23}$Na, with I=3/2, a single NQR line is
expected per Na site, and indeed we detected three sites with narrow NQR
linewidths. (b) For $^{59}$Co, with nuclear spin $I=7/2$ we display here the
four distinct lines corresponding to the higher frequency $(7/2\rightarrow
5/2)$ transitions.}
\label{FigNQR}
\end{figure}

We found that the simplest NMR spectra among the high sodium content phases
\cite{ourEPL2008} were observed for $x=0.67$. $^{23}$Na NMR only exhibited
three distinct Na sites on powder samples aligned within a polymer matrix by
an applied field \cite{NaPaper}. In such samples the crystallite grains were
aligned along their $c$ direction while their $a,b$ directions are at random
in the perpendicular plane {(the alignment procedure was recalled in
ref.~[18] of our ref.~\cite{CoPaper})}. The Na lines were not fully resolved,
which could have been attributed to intrinsic sample disorder. In NQR, no
magnetic field being applied, the spectra do not depend on the grain
orientations. So the detection, in the same sample, of three well resolved
narrow NQR lines (fig.~\ref{FigNQR}a) with the expected quadrupolar
frequencies $\nu _{Q}$, convinces us that the NMR resolution was limited by
a small distribution of orientations of the crystallites.

\begin{figure}[tbp]
\onefigure[width=1\linewidth]{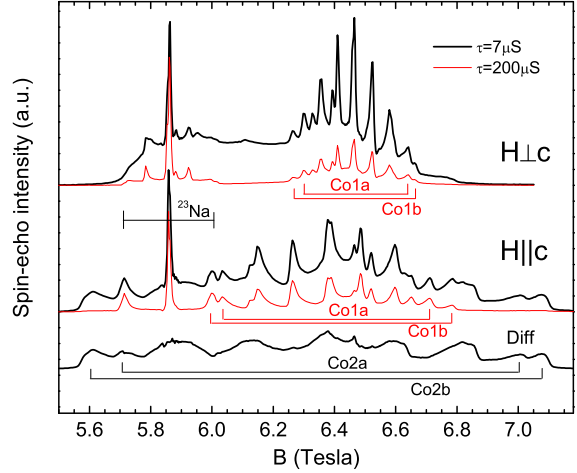}
\caption{Asymmetry of the $^{59}$Co NMR spectra. $^{59}$Co NMR spectra
obtained by sweeping the applied field at fixed frequency and monitoring the
NMR signal by the magnitude of the spin echo signal. The top spectra are
taken with the external field applied parallel to the ab plane, while the
lower ones are taken with $H\parallel c$. The black spectra taken with short
delay between pulses ($\tau =7\mu sec$) give the overall NMR
signal including all sites. The red spectra of the non magnetic Co1 sites,
which have longer transverse spin-spin relaxation time, are separated out
using a longer delay ($\tau =200\mu sec$)\ Scaling it to
deduce it from the full spectrum (as in ref.~\cite{CoPaper})
allows to deduce the signal of the Co2 sites displayed as Diff. in the lower
panel for $H\parallel c$. We have marked the positions of the
$(5/2\rightarrow 7/2)$ outer quadrupolar transitions for the Co1 and Co2
spectra for $H\parallel c$. For $H\perp c$ those are still well resolved for
the Co1 with an extension reduced by a factor 2 (as marked). This is not the
case for the Co2 signals for which $\eta \neq 0$.}
\label{FigNMR}
\end{figure}

For $^{59}$Co, we detected simple NQR spectra as well, with \textit{four
well resolved sites} (see fig.~\ref{FigNQR}b), grouped two by two, Co1a and
Co1b with lower NQR frequencies than Co2a and Co2b. The overlap between the
NMR spectra of these Co sites results in complicated NMR spectra as shown
for instance in fig.~\ref{FigNMR}. While the NMR differentiation techniques
used in ref.~\cite{NaPaper} allowed us to separate there up to five $^{59}$%
Co NMR signals, one of them was found to result from a small fraction of
polycrystalline grains of the sample which did not align in the field
\footnote{Comparing different batches of samples, we noticed that the part of the
spectrum corresponding to a large NMR shift for both directions of the field
was sample dependent. It was erroneously assigned in the first experiments
of ref.~\cite{NaPaper} to a site labelled Co3 with an \textquotedblleft
isotropic\textquotedblright shift. The analysis of the NMR shift data for
the four Co sites will be detailed in a forthcoming publication (ref.~\cite%
{IrekCo}). The spectra shown in fig.~\ref{FigNMR} on a better oriented
sample only retain the four site signals, in agreement with NQR.}.

The fractional occupancies of the various sites could be better fixed by
comparing first the signal intensities of the sites with neighboring
quadrupole frequencies. Similar ratios Co1a/Co1b=1.95(0.1) and
Co2b/Co2a=1.9(0.2) were obtained from NQR data of fig.~\ref{FigNQR}a, after
correcting for the spin-spin $T_{2}$ decay and for the $\nu ^{2}$ frequency
dependence of the signal intensity. The $\nu _{Q}$ values obtained by NQR
allowed us to reduce as well the input parameters in the analysis of the $%
^{59}$Co NMR spectra and to confirm this estimate of Co1a/Co1b =1.85(0.2)
and that Co2b has definitely a larger intensity than Co2a. To complete these
comparisons one needed to determine the relative intensities of the Co1 and
Co2 sites. The best accuracy has been obtained from a comparison of the
total intensity of the Co1 sites signal to that of the full $^{59}$Co NMR
spectrum, which gave  Co1/(Co1+Co2)=0.26(0.04). This allows to obtain the
fractional occupancies given in table~\ref{Table1}, which were quite helpful
in the identification of the structure.

\begin{table}[tbph]
\caption{EFG parameters $\nu _{Q}$\ and $\eta $ for the
different sites and relative intensities $I_{exp}$ deduced from NQR or NMR
(see text). $I_{exp}$ can be compared to the intensities $I_{str}$ expected
from the site occupancies $N_{s}$ of the structure depicted in fig.~\ref{FigStructure}.}
\label{Table1}
\begin{center}
\begin{tabular}{|c|c|c|c|c|c|}
\hline
\raisebox{-1.50ex}[0cm][0cm]{Site} & \multicolumn{3}{|c|}{Experiment} &
\multicolumn{2}{|c|}{Structure} \\ \cline{2-6}
& \textit{$\nu $}$_{Q}$ (MHz) & $\eta $ & $I_{exp}(\%)$ & {$N_{s}$} & {$%
I_{str}(\%)$} \\ \hline
Na1 & 1.645(5) & $<0.01$ & 30(5) & 2 & 25 \\ \hline
Na2a & 1.74(1) & 0.84(2) & 33(5) & 3 & 37.5 \\ \hline
Na2b & 1.87(1) & 0.89(2) & 37(5) & 3 & 37.5 \\ \hline
Co1a & 1.193(2) & $<$0.01 & 17(3) & 2 & 16.7 \\ \hline
Co1b & 1.392(2) & $<$0.01 & 9(2) & 1 & 8.3 \\ \hline
Co2a & 2.187(3) & 0.36(1) & 25(3) & 3 & 25 \\ \hline
Co2b & 2.542(3) & 0.36(1) & 49(4) & 6 & 50 \\ \hline
\end{tabular}%
\end{center}
\end{table}

\section{Lattice structure and site determination}

\begin{figure*}[tp]
\begin{center}
\onefigure[width=0.75\linewidth]{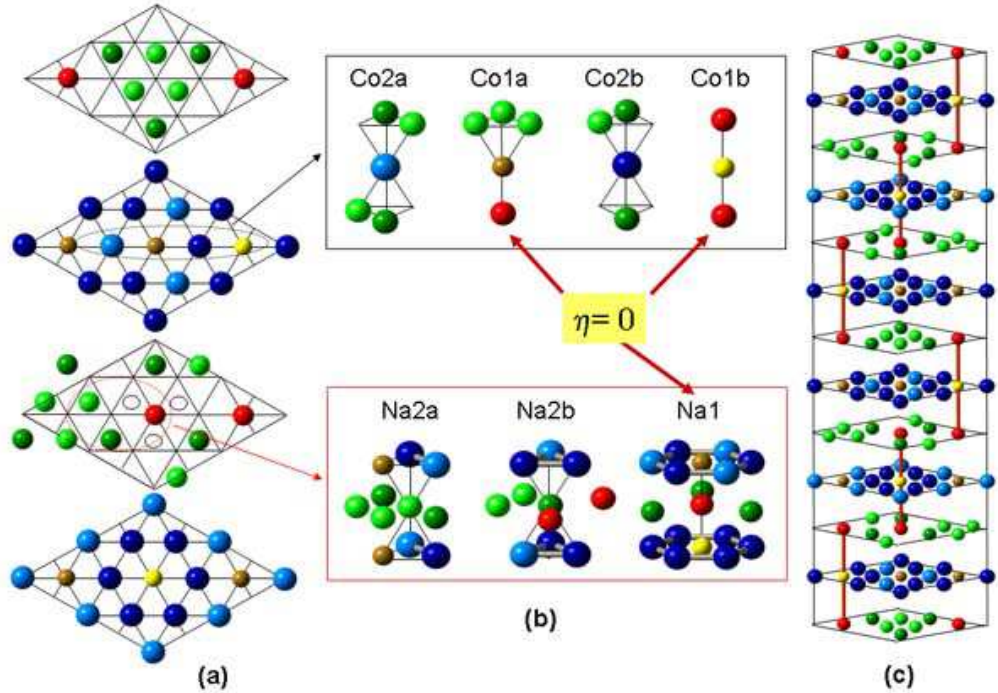}
\end{center}
\caption{Three dimensional structure of Na$_{2/3}$CoO$_{2}$. \textbf{(a)}
The two dimensional Na unit cell proposed in ref.~\cite{Hinuma} is
represented in the top layer with respect to the 12 initially
indistinguishable Co atoms of the underlying triangular layer (the oxygen
atoms layers above and below the cobalt layer are omitted). It contains two
Na1 sites (red) sitting on top of Co and six Na2 sites, topping a triangle
of Co, with two distinct Na planar environments (light and dark green). The
3D stacking which fits our data involves a lower Na layer shifted by -2%
\textbf{\textit{a}}. As the Na2 sites of the two layers should project at
the centers of distinct Co triangles, \textbf{\textit{a}} axis mirroring of
the Na pattern is required. This planar Na structure can be considered as an
ordered pattern of di-vacancies in the Na2 plane, which displace a third Na2
on an Na1 site in a three fold symmetric environment (one such di-vacancy is
illustrated in the lower Na plane, where circles mark the three missing Na2
sites). \textbf{(b)} In the upper panel, the Na environments of the four Co
sites differentiated in this structure are displayed (the smaller size of
the orange and yellow balls is meant to distinguish Co sites with Na1
neighbors). In the lower panel, the 3 Na sites are represented with their Na
and Co environments deduced from the stacking of cobalt planes shown in
\textbf{(a)}. Three fold rotation symmetry is seen for Co1a , Co1b and Na1
for which $\eta $=0. \textbf{(c)} The four layers of \textbf{(a)}
are represented at the top and the 3D stacking of the following Na layers is
pursued downwards by -2\textbf{\textit{a}} translations and mirroring, which
shifts the Co1b site by -2\textbf{\textit{a}} between Co layers (red bars).
The unit cell is completed after six shifts.}
\label{FigStructure}
\end{figure*}

Na ordering patterns have been suggested both from diffraction experiments
\cite{Roger} and calculations restricted to electrostatic interactions
between ions at fixed positions \cite{Roger,Zhang}. We found that the type
of Na ordering proposed for $x\geq 0.75$, which consists in an ordering of
droplets of Na2 tri-vacancies in which Na1 trimers are stabilized \cite%
{Roger}, cannot explain the present data (see fig.~\ref{FigStructure} for
the usual convention taken for Na1 and Na2 sites). Such calculations are
somewhat inconclusive on the possible stable structure for $x=2/3$. A recent
quantum chemistry calculation allowing the relaxation of the site positions
\cite{Hinuma} suggest for this Na content a relatively simple stable
structure for $x=2/3$, with a 2D unit cell containing 12 cobalt atoms. It
corresponds to 2D ordering of Na2 di-vacancies which stabilize a single Na1
site at their center in a position with axial symmetry. This unit cell which
is shown on top of fig.~\ref{FigStructure}a contains 8 Na for 12 Co atoms in
the Co layer.

We shall see that we could fit perfectly NMR/NQR data by choosing the
stacking pattern of this unit cell shown in fig.~\ref{FigStructure}a, which
is in good agreement with the "rules" suggested in ref.~\cite{Hinuma}, $i.e.$
minimize the number of Co sites with two Na1 above and below. The
differentiation of sites intrinsic to this structure gives 4 Co and 3 Na
sites, which are distinguished in fig.~\ref{FigStructure}a and fig.~\ref%
{FigStructure}b. We find that six layers of CoO$_{2}$ are required to
complete the 3D unit cell shown in fig.~\ref{FigStructure}c. This
corresponds to a lattice parameter $3c$, where $c$ is the usual unit cell
length involving two CoO$_{2}$ layers in the basic hexagonal structure, in
agreement with the observation done on this phase by X rays \cite{ourEPL2008}%
. Although this unit cell contains 72 cobalt, 144 oxygen and 48 sodium
atoms, the local site differentiation of fig.~\ref{FigStructure}a,b is
maintained in all the 3D crystal structure, with the in plane multiplicities
$N_{s}$ of the 4 Co and 3 Na sites given in table~\ref{Table1}. The
agreement found there with the relative intensities of the corresponding
NMR/NQR lines allows an unambiguous assignment of the different lines to the
sites as given in fig.~\ref{FigStructure}b. The Co1a line which has the
lower intensity is naturally assigned to the Co site of the 2D unit cell
which has Na1 sites above and below, while Co1b corresponds to the two
cobalt with one Na1 on one side and three Na2 on the other side (fig.~\ref%
{FigStructure}b). The two other sites with multiplicity 3 and 6 are then
naturally assigned to Co2a and Co2b, while the relative intensity of
0.26(0.04) found for the Co1 spectrum agrees perfectly with the expected
3:12 ratio for the structure of fig.~\ref{FigStructure}a.

As for the $^{23}$Na NMR data \cite{NaPaper}, the lower intensity line
corresponds then to the Na1 sites, while the two others with similar
intensities can be indifferently assigned to Na2a and Na2b. Let us point out
that Chou \textit{et al.} \cite{Chou2} reported, on a single crystal, a well
resolved $^{23}$Na spectrum taken at 20K that we could ascribe without
ambiguity to the same H67 phase, from the $^{23}$Na NMR shifts and
quadrupolar splitting values. They attempted to assign this three line
spectrum to a phase with $x=0.71$, as obtained by the chemical analysis of
their sample.\ However the relative intensities reported, 26.5{\%}, 36.3{\%}%
, 37.1{\%}, correspond extremely well within their 3{\%} accuracy to the
2:3:3 ratio for the Na sites, which independently supports the validity of
the structure proposed here.

\section{Symmetry of the sites}

An important cross-test of the validity of this structure is the symmetry of
the various sites, which could be probed through that of the EFG\ tensor.\
The NQR frequencies are indeed determined by the principal values of the EFG
tensor, which is diagonal in a frame $(X,Y,Z)$ linked with the
crystallographic structure $(|V_{ZZ}|\geq |V_{YY}|\geq |V_{XX}|)$. Laplace
equation reduces those into two parameters, the quadrupolar frequency%
\begin{equation}
\nu _{Q}=3eQV_{ZZ}/[2I(2I-1)]
\end{equation}%
and the asymmetry parameter%
\begin{equation}
\eta =(V_{XX}-V_{YY})/V_{ZZ},
\end{equation}%
which vanishes identically for sites with axial symmetry.

From fig.~\ref{FigStructure}b it is easy to see that this is the case for
the Na1 site as well as the Co1a and Co1b sites, for which $c$\textbf{\ }is
a threefold symmetry axis in the proposed structure. In NMR the weaker
intensity Na site had been found \cite{NaPaper} with $\eta =0.01(1)$,
confirming the identification done with the Na1 site, while the Na2a and
Na2b have large $\eta $ values in agreement with the non axial structure.
Similarly, one can see in fig.~\ref{FigNMR} that the quadrupole splitting of
the $^{59}$Co NMR of the Co1 sites is reduced by a factor two in the $ab$
direction with respect to that in the $c$ direction. This is expected for
sites with a threefold or fourfold symmetry axis for which $%
V_{XX}=V_{YY}=-V_{ZZ}/2$, which corresponds to an axial symmetry for the
EFG. In such a case the quadrupole splitting is identical for all field
directions in the $ab$ plane, so that the NMR spectra are still well
resolved for $H\perp c$, even for a powder distribution of the $ab$
orientations, as seen in fig.~\ref{FigNMR}. On the contrary one can see that
the $H\perp c$ spectra of the Co2 sites are less resolved and have a field
extension comparable with that in the $H\parallel c$ direction, implying $%
\eta \neq 0$. A full study of the NMR and NQR spectra to be published
elsewhere allowed us to deduce $\eta =0.36(1)$ for both Co2 sites.

So we do find that all the sites with axial symmetry of fig.~\ref%
{FigStructure}b have been properly identified, \textit{which secures then
the validity of the proposed structure}. This confirms our $x=0.67(1)$
estimate for the Na content of this phase and establishes the validity of
the relationship between $c$ axis parameter and Na content reported in \cite%
{ourEPL2008,Lang2}. {We believe that other calibrations \cite{Shu,Chou2}
based on \textit{chemical} analyses of samples overestimate the actual
content of the majority phase from which Na tends to be expelled. This is
particularly true for single crystal samples synthesized with a large excess
of Na , while for our single phase ceramic samples reacted in solid state,
Na excess with respect to nominal content is unlikely.}

\section{Magnetic properties of the Co sites}

\begin{figure}[tbp]
\onefigure[width=1\linewidth]{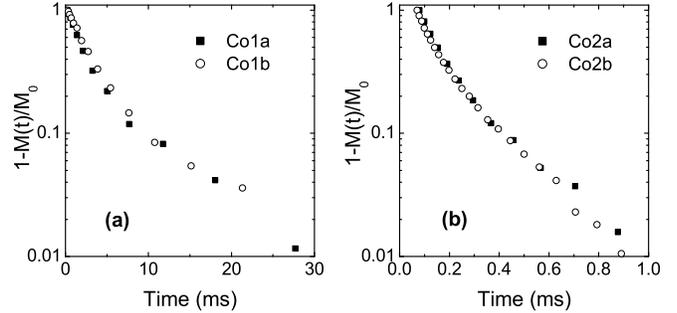}
\caption{Spin lattice relaxation of the Co sites taken at 4.2K. The
magnetization recoveries of the $^{59}$Co NQR signals are seen in (a) to be
identical for the lower frequency signals of Co1a and Co1b sites, and
similarly in (b) for the higher frequency signals of Co2a and Co2b. The
relaxation rate is about 30 times larger for the latter,
which differentiates markedly the magnetic Co2 sites from the non magnetic
Co1 sites.}
\label{FigT1}
\end{figure}

It has been shown for long that this particular phase has an anomalous Curie
Weiss behavior extending down to the lowest temperatures \cite{NaPaper} and
a resistivity which only reaches a $T^{2}$ dependence below 2~K \cite{LiSY}.
 The local magnetic parameters could be studied through NMR shift data \cite%
{CoPaper} or spin lattice relaxation $T_{1}$ measurements which probe
directly the dynamic susceptibilities. Here, to compare in detail the
magnetic properties of the four cobalt sites we took advantage of the fact
that the NQR lines are fully resolved spectrally in fig.~\ref{FigNQR},
contrary to the NMR spectra of fig.~\ref{FigNMR} which largely overlap. This
allowed us to probe independently the $T_{1}$ values of the four sites
through monitoring the recovery of each NQR signal after saturation. The
typical data taken at 4.2~K shown in fig.~\ref{FigT1} immediately allow us
to evidence that the Co2a and Co2b sites have identical $T_{1}$ values,
\textit{which are about 30 times shorter} than that of the non magnetic Co1a
and Co1b sites. For the latter the $T_{1}$ values are identical as well as
the NMR shifts, which have an isotropic orbital component, which allowed us
to evidence that they correspond to non magnetic Co$^{3+}$ sites \cite%
{CoPaper,ourEPL2008}. This illustrates that the magnetic properties are
sustained by the Co2 sites which sense directly the on site magnetism that
yields much larger spin lattice relaxation rates (and NMR shifts \cite%
{CoPaper}). On the contrary, the Co1 and Na sense the magnetism of the Co2
sites only through transferred hyperfine couplings. For instance the fact
that Na2a and Na2b  display smaller NMR shifts than Na1 \cite{NaPaper} can
be assigned to the larger number of magnetic Co2 near neighbours for the
latter (fig.~\ref{FigStructure}b).

More importantly the present results prove that Co2a and Co2b sites,
although quite distinguished by their EFG values induced by the local
structure, \textit{are quite identical on their electronic and magnetic
properties}. So the Na organization above and below the Co plane allows to
pin Co$^{3+}$ ionic states on the Co1 sites, by lowering their energy levels
so that holes do only occur on the Co2 sites. For the understanding of the
electronic properties one \textit{should then consider only two types of
electronic sites} (3 Co1 and 9 Co2) in a single plane unit cell as shown in
fig.~\ref{FigKagome}. There one can notice that the Co2 sites\textit{\ are
quite remarkably organized on a kagom\'{e}} structure while the Co$^{3+}$
sites form the complementary triangular lattice. The fractional occupancy
1:4\ of the Co1 sites agrees well with the previous estimates of the Co$%
^{3+} $concentration, which allows us to establish that the partly filled
Co2 sites bear then a hole concentration of 4/9 per site, $i.e.$ an
effective charge Co$^{3.44+}$, similar to that deduced before \cite%
{CoPaper,ourEPL2008}, and far less than expected for a Co$^{3+}$/ Co$^{4+}$
scenario.

{Let us point out that a similar ordered charge disproportionation (OCD) has
been found in the case of nickelates \cite{Nickelates}, but with opposite
magnetic effects, i.e. local moments on one site out of three and charge
delocalisation on the others. The experimental situation here is rather
analogous to that found in the cubic CsC$_{60}$ compounds where local
singlets are favoured on some C$_{60}$ balls while electrons delocalize on
the remaining ones \cite{CsC60}.}

\section{Discussion}

\begin{figure}[tbp]
\onefigure[width=1\linewidth]{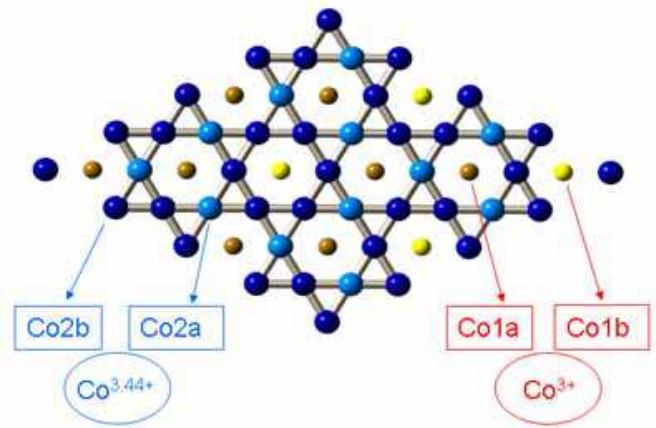}
\caption{Two dimensional charge distribution in the Co planes of Na$_{2/3}$%
CoO$_{2}$. The resulting 2D structure of the Co planes corresponds to a
perfect Co2 kagom\'{e} lattice (fig.~\ref{FigStructure}c), once the
minor difference between the electronic properties of Co2a and Co2b sites is
neglected. The Co1 sites constitute the complementary triangular lattice.}
\label{FigKagome}
\end{figure}

{So far, most electrostatic calculations do give indications on possible Na
orderings, but they do not introduce the electronic energies sufficiently
well to give any insight on the electronic order associated. The calculation
of ref.~\cite{Hinuma} which takes into account the electronic band energy
(with a variant of the Local Density Approximation called GGA) yields an
atomic structure in agreement with that proposed here. This underlines the
importance of the electronic energy in pinning the Na order. Although the
authors establish that the same Na structure is stable if a large on site U
is introduced in the GGA, nothing in the reported results helps so far to
establish the stability of the OCD of fig.~\ref{FigKagome}. The competition
between Hund's energy and band energy which stabilizes local moments in the
case of nickelates \cite{Nickelates} might rather favour the filling of
selected Co$^{3+}$ sites in the present case. It would certainly be helpful
to extend then the GGA calculations to check whether this approach is
sufficient to explain a decrease of the ground sate energy by the OCD, or
whether one has to resort to a more refined treatment of the electronic
correlations. The derived band structure would then allow comparisons with
the hyperfine and EFG parameters determined from our NMR data. Depending
which bands ($a_{1g}$ and/or $e_{g}^{^{\prime }}$) participate to the Fermi
surface in the reduced Brillouin zone, might allow one to interpret the
small pockets detected by transport experiments \cite{Balicas}. It might
also be worth to attempt ARPES experiments \cite{Yang} on ordered Na phases
on the external layer of single crystals to detect their imprint on the
Fermi surface, although this appears an experimental challenge.}

On theoretical grounds, one would like as well to understand whether the
Kagom\'{e} structure has any specific role in the physics of cobaltates, as
had been underlined by Koshibae and Maekawa \cite{Koshibae}. These authors
anticipated that the directionality of the transfer integrals between Co
sites in the triangular lattice favors electronic \textit{wave functions}
restricted to orbitals organized on a kagom\'{e} lattice. In a uniformly
charged case four such interpenetrating lattices are degenerate, but the Na
self organization might merely select here one of these kagom\'{e} lattices?

As for the magnetic properties of the cobaltates, the experimental evidence
given here that metallicity and magnetism are combined in the correlated
electron kagom\'{e} structure is at odds with most proposals concerning the
electronic structure of cobaltates \cite{Baskaran,Bernhardt,Chou,Marianetti}%
. It has indeed most often been anticipated that Na2 vacancies pin localized
spins which would dominate the Curie Weiss magnetic properties. In these
approaches, close to the ionic Co$^{3+}$/ Co$^{4+}$ scenario, the metallic
properties are associated with quasi non magnetic Co$^{3+}$ sites assumed to
retain Na configurations similar to that of Na$_{1}$CoO$_{2}$. On the
contrary, our results establish that the Na1 sites are linked to non
magnetic Co$^{3+}$on the Co1, and that metallicity occurs on the Co2 sites
which, as viewed in fig.~\ref{FigStructure}b, have a low Na coordinance (3
or 4) compared to the 6 fold one of Na$_{1}$CoO$_{2}$.

So the present results give a simple realistic structure on which approaches
taking better into account the electronic correlations, such as DMFT \cite%
{Marianetti} could be applied.\ It might even be used as a benchmark to
develop and test the capabilities of electronic structure calculations of
correlated electron systems.This could possibly permit to understand
altogether why the dominant electronic correlations are ferromagnetic for $%
x\geq 2/3$ \cite{ourEPL2008} and AF below \cite{Lang2}. Also with this well
characterized 3D order one should understand whether specific inter-plane
exchange paths could justify the absence for $x=2/3$ of the A type AF 3D
order \cite{Bayrakci,Helme} found for $x>0.75$. Conversely one might
wonder whether this difference can be explained by an incidence of the 2D
frustration inherent to the kagom\'{e} structure, establishing for instance
an underlying spin liquid component as found in the $S=1/2$ kagom\'{e}
lattice \cite{Mendels3}? The present results not only raise these questions
but open a path to solve them, as the experimental approach highlighted
here, which combines the structural determination with its local impact on
magnetism can be extended to the diverse cobaltate phases with distinct
ground state properties.

\section{Aknowledgments}

We acknowledge J.~Bobroff, G.~Collin, G.~Lang, P.~Mendels, M.~Rozenberg and
F.~Rullier-Albenque for their constant interest, for helpful discussions and
comments concerning the manuscript. We acknowledge financial support by the
ANR (NT05-441913) in France. Expenses in Orsay for A.~D. and I.~M. have been supported by the
\textquotedblleft Triangle de la Physique\textquotedblright. T.~Platova
has obtained a fellowship from the  E.U. Marie Curie program
\textquotedblleft Emergentcondmatphys\textquotedblright\ for  part of her
PhD work performed in Orsay.

\end{document}